# Uncovering multifunctional mechano-intelligence in and through phononic metastructures harnessing physical reservoir computing

Yuning Zhang, Aditya Deshmukh, K. W. Wang*

Department of Mechanical Engineering, University of Michigan, Ann Arbor, MI 48109, USA

*Corresponding author: K. W. Wang

Email:  kwwang@umich.edu

**Abstract**

The recent advances in autonomous systems have prompted a strong demand for the next generation of adaptive structures and materials to possess more built-in intelligence in their mechanical domain, the so-called mechano-intelligence (MI). Previous MI attempts mainly focused on specific designs and case studies to realize limited aspects of MI, and there is a lack of a systematic foundation in constructing and integrating the different elements of intelligence in an effective and efficient manner. Here, we propose a new approach to create the needed foundation in realizing integrated multifunctional MI via a physical reservoir computing (PRC) framework. That is, to concurrently embody computing power and the various elements of intelligence, namely perception, decision-making, and commanding, directly in the mechanical domain, advancing from conventional adaptive structures that rely solely on add-on digital computers and massive electronics to achieve intelligence. As an exemplar platform, we construct a mechanically intelligent phononic metastructure with the integrated elements of MI by harnessing the PRC power hidden in their high-degree-of-freedom nonlinear dynamics. Through analyses and experimental investigations, we uncover multiple adaptive structural functions ranging from self-tuning wave controls to wave-based logic gates. This research will provide the basis for creating future new structures that would greatly surpass the state of the art — such as lower power consumption, more direct interactions, and much better survivability in harsh environment or under cyberattacks. Moreover, it will enable the addition of new functions and autonomy to systems without overburdening the onboard computers.

**Significance Statement**

We uncover a new approach to create and embody mechano-intelligence in structures based on the framework of physical reservoir computing and investigate the concept in and through a mechanically intelligent phononic metastructure. This idea builds the much needed foundation in systematically integrating the various elements of intelligence, including perception, decision-making, and commanding, mainly in the mechanical domain to achieve broad impact in future systems, such as automated robots and vehicles, smart wearable devices and intelligent infrastructures. This new class of adaptive structures revolutionizes current practices that solely rely on add-on computers to achieve intelligent functions. It is transformative in advancing the state of the art with lower power consumption, more direct interaction, better cybersecurity, and greater survivability in harsh environments.

**Main Text**

**Introduction**

The prominent demand for increasing level of autonomy and adaptation in high-performance engineering systems, such as smart robotics, automated vehicles, and adaptive infrastructures, has led to the emergence of mechanically intelligent structural and material systems. These systems ideally would possess the ability to perceive information from sensory signals, make decisions, and command and respond in the mechanical domain, enabling them to adapt more directly to changing environments and mission requirements with higher efficiency, resiliency, and security. Previous studies have attempted to address certain aspects of intelligence into the design of various systems [1]–[17]. While promising, these mechano-intelligence (MI) efforts have been primarily limited to specific cases with narrowly-defined functions. Overall, a systematic foundation remains lacking for effective synthesizing and integrating the various intelligent elements, namely information perception, decision-making, and commanding, in multifunctional structures.

Inspired by the remarkable success of artificial neural networks, recent advancements have unlocked the potential of physical reservoir computing (PRC) [18], [19], i.e., implementing reservoir computing in a physical body-based recurrent neural networks leveraging their high-degree-of-freedom and nonlinear dynamics. Reservoir computing employs a fixed-interconnection network as the reservoir and only requires training for the readout layer, resulting in low training cost. Moreover, due to the fixed nature, their physical implementation becomes very promising, which opens up the excellent possibility of using mechanical systems to conduct computing [3], [20]–[24], ideal for our proposed MI. However, despite the great potential of the PRC framework, the efforts have mostly focused on computation only, and not yet been utilized to construct MI to enable engineering-relevant functionalities.

Based on the above background, the vision of this research is to advance MI by harnessing PRC as the needed foundational framework to integrate the essential elements of intelligence in the mechanical domain at the system level, such as perceive information from sensory processing, make decisions, and command actions to actuators, and eventually form the basis for future adaptive structures with embodied multi-faceted functional-relevant intelligence. That is, by training these structures to learn complex input-output mappings through a trainable linearly weighted readout layer, they can exhibit versatile and multi-tasking information-processing capability, which is crucial for the various intelligent elements and can naturally serve as a promising foundation for their integration. To bridge the technological gaps, this work combines experimental investigation, theoretical analysis, and numerical simulation to explore a mechano-intelligent phononic metastructure (PM) with integrated perception, decision-making, and commanding capabilities, achieved by harnessing its dynamics as a physical reservoir. As typical engineered matter, PMs have emerged as the building blocks for numerous advanced systems, exhibiting significant advantages over conventional materials by enabling nontraditional wave characteristics [25]–[31]. Of particular significance is the formulation of bandgaps due to Bragg's scattering or local resonance effects — a frequency region where the incident wave propagation is prohibited, which has facilitated a wide range of engineering applications, spanning from vibration and noise controls, structural monitoring, to energy harvesting [32]–[35]. In this study, we harness the PRC-enabled MI for multiple functionalities in wave transmission controls as well as wave-based logic gates. Through the PRC framework, we design a highly tunable one-dimensional PM that can effectively adjust its wave characteristics in response to different input excitations by perceiving sensory signals, making informed decisions, and commanding actions to alter its configuration. Our experimental results demonstrate that the same PM can effectively perform multiple intelligent functional tasks, such as self-adaptive frequency-selective wave attenuation and amplitude-selective wave filtering, without prior knowledge of input signals or the need for complex digital controllers often employed in mechatronics-based systems. In other words, the perception, decision-making, and commanding capabilities are embedded in the mechanical metastructure for integrated and multifunctional MI. Moreover, we go beyond wave transmission control and explore new potentials in achieving wave-based logic gates by abstracting the binary logical bits based on the wave transmission characteristics. We successfully realize all the basic logic gates via the proposed self-adaptive bandgap tuning mechanism, thereby offering a new paradigm for phononic wave-information processing and communication systems. Unlike conventional adaptive structures solely utilize add-on digital computers to enable intelligence, the built-in MI can create functionalities that are more energy-efficient, more direct in interaction, more durable in extreme environments, and less vulnerable against cyberattacks.

**Results**

**Phononic metastructure reservoir architecture**. The proposed PM framework, as depicted in Fig. 1A and Fig. 1D, consists of a chain of 11 periodic diatomic modules. As highlighted in Fig. 1C, each metastructure module comprises two rigid masses made of cast acrylic sheets coupled with a pair of bent spring steel beams (type I), while the modules are inter-connected via another pair of bent beams (type II). The PM is fixed at the left end to a 3D-printed support, and its right end is connected to a linear motion actuator capable of generating a static displacement and altering the overall length of the PM, thus modifying its dynamic properties. We consider the longitudinal elastic wave propagation in the PM along $+x$ direction under harmonic excitation. The input wave source is generated by applying harmonic magnetic forces to the leftmost module using an electromagnet coil shaker powered by a waveform generator. To measure the wave dynamics along the PM chain, a scanning laser Doppler vibrometer is employed to obtain the out-of-plane ($y$ direction) velocity at the mid-points of the bent beams, resulting from the longitudinal vibration of the rigid masses (See SI Appendix, Section S1 for fabrication details and experimental setups).

Figure 1A and Fig. 1B illustrates the proposed PM reservoir architecture to enable intelligent wave functionality adaptation. The key concept is to treat the PM as a physical neural network, i.e., the metastructure reservoir layer, and harness its wave dynamics as informative computational resource for computing the execution command of the global displacement $X_a$. This command is subsequentially fed to the linear actuator to deform the PM and modify its wave transmission characteristics according to specific mission requirements, such as blocking or passing the incoming wave. The reservoir output is obtained by a linearly weighted readout layer downstream to the metastructure reservoir layer. Here, the steady-state velocity amplitudes $A_1, A_2, \ldots, A_N$ of selected bent beams are collected as the readouts, and the final output is determined through a linear combination of the readouts, readout weights $w_1, w_2, \ldots, w_N$ and bias $b$, i.e., $X_a = \sum_i w_i \cdot A_i + b$, where the weights and bias are the trainable parameters that require tuning for different functional tasks. With this arrangement, the PM reservoir can be trained to learn the complex mappings between the input excitation features (e.g., input amplitude $A_0$ and frequency $f$) and the desired actuator displacement $X_a$. Consequently, as a physical neural network, the PM can intelligently adapt to different input waves by perceiving information from sensory processing, making decisions on reconfiguration, and commanding the corresponding actuations.

In order to design a targeted input-output mapping for controlling the wave propagation, we first investigate the tunable wave transmission characteristics of the PM under different input frequencies and global configurations (deformations) using experimental, numerical and theoretical tools. In experiments, the PM is uniformly configured to 20 global lengths $X_a$ from the initially undeformed case $X_a = 0$ cm to the maximum stretch $X_a = 12.5$ cm. For each configuration, a scanning frequency signal with a voltage amplitude $v_0 = 1\ V$ is applied to the electromagnetic coil shaker, and the frequency response of each module is obtained via fast Fourier transformation.

In addition to the experimental investigations, an analytical model is established by modelling each pair of bent beams as a massless and damped nonlinear spring connecting two lumped masses $m_1$ and $m_2$, as shown in the inset of Fig. 1C. The restoring forces of the two types of bent beam pairs under various deformations are mathematically formulated and compared to the experimentally measured quasi-static behavior of the PM modules in Fig. 2A, which finds excellent agreement (see SI Appendix, Section S2 for details on the mathematical modeling of the PM modules). A highly nonlinear and stretch-hardening behavior of the PM modules is observed, which offers the means to tune the local stiffness of the PM through global length reconfiguration, thereby altering its wave physics.

With the analytical model, the wave dynamics of the PM is then numerically studied by solving its governing equations (see SI Appendix, Section S2 for details on the governing equations). In our simulations, a small-amplitude input force excitation $F_{in} = A_0\ sin\ (2\pi ft)$, where $A_0 = 0.05\ N$, is applied to the leftmost module, which ensures that the masses undergo only small-amplitude vibrations around their equilibrium positions for the different global length configurations. To gain deeper physical insights, a linear dispersion analysis is conducted to obtain the theoretical band structure of the PM, i.e., the theoretical solutions of the dispersion relationship $\omega = \omega(k)$ between wave number $k$ and angular frequency $\omega$, by assuming an undamped and infinite chain of periodic modules (see SI Appendix, Section S3 for details on the linear dispersion analysis).

To quantify the wave propagation characteristics, we introduce the transmission ratio (TR), which is defined as the ratio between the average velocity amplitude of the output modules (the two right modules) and that of the input modules (the two left modules). Figure 2B presents the experimentally measured TR spectrum of the PM with respect to the input frequencies under different global length configurations, which clearly demonstrates the existence of experimental passbands and stopbands. Within the passbands, high TRs ($\cong 1$) are observed, implying that most of the wave energy propagates to the right side of the PM, while the TRs drop significantly below 0.1 within the stopbands, indicating the blocking of incoming waves. Furthermore, the theoretically predicted bandgap boundaries are also plotted in this figure, which closely match the experimental results. Notably, we observe an increase in boundary frequencies due to the PM module hardening effects under higher global stretch, highlighting the highly tunable wave physics of the PM. Fig. 2C provides a comparison of the experimental and numerical TRs, along with the theoretical band structures, for two specific configurations, i.e., $X_a = 0$ cm and $X_a = 7.3$ cm. The agreement among the results from experiments, numerical simulations, and theoretical predictions confirms the accuracy of the analytical model.

**Self-adaptive wave transmission control**. We can then manipulate the tunable band structure of the PM to design targeted mappings $X_T = T(f, A_0)$ between the targeted actuator displacement $X_T$ and the input wave features, including input frequencies $f$ and amplitudes $A_0$, to achieve self-adaptive wave transmission control. With these mappings, the PM reservoir will execute adaptive actions based on the targets to pass or block incoming waves in response to different excitations. Firstly, we focus on frequency-selective wave transmission control by training the PM reservoir to learn a frequency-dependent mapping $X_T = T(f)$ with fixed input amplitude. For example, Figure 3A illustrates a self-adaptive wave blocking task to prohibit the wave to propagate for any frequency by setting the targeted actuator displacement located within the stopband region. In our study, the PM is always excited from the initially undeformed configuration of $X_a = 0$ cm, as represented by the green dashed line. We are interested in the operating frequency range of $80\text{Hz}$ to $136\text{Hz}$, where the PM initially exhibits a stopband between $80\text{Hz}$ and $114\text{Hz}$ and a passband between $114\text{Hz}$ and $136\text{Hz}$. The targeted mapping, indicated by the black dots in Fig. 3A, is designed to be within the theoretical bandgap, promoting the PM to reconfigure itself and block the wave. Specifically, for input frequencies within the initial stopband, the targeted displacements are set to zero as the wave is already blocked and no action is required. However, for input frequencies within the initial passband, finite targeted displacements are expected to shift the band structure and prevent wave propagation (see SI Appendix, Section S4 for details on the targeted mapping design). As a result, the PM can adaptively respond to different input frequencies and make frequency-based decisions to block the wave.

To train the PM reservoir, we collect the experimental wave dynamics data for 128 different input frequencies randomly selected from $80\text{Hz}$ to $136\text{Hz}$ as the training sample. During the training phase, we optimize the weights of the readout layer $w_1, w_2, \ldots, w_N$ and the bias b to minimize the error between the targeted displacements $X_T$ and the PM reservoir outputs $X_a = \sum_i w_i \cdot A_i + b$ across all the training samples. The normalized mean square error is employed to measure the error between the targets and the reservoir predictions, and the optimal readout weights can be obtained through a least square regression. To evaluate the performance of the PM reservoir, we randomly selected 32 unseen frequencies from the specified frequency range for testing purpose (see SI Appendix, Section S4 for details on the PM reservoir training and testing). The training and testing results for the self-adaptive wave blocking task are depicted in Fig. 3A, left, where the red and blue dots represent the reservoir outputs on training and testing samples, respectively. It is observed that despite some deviation from the targets, all the training and testing samples end up within the bandgap region, indicating the successful implement of the wave blocking task. We also measure the TRs for all the training and testing frequencies under the initially undeformed position and the final reconfigured position (Fig. 3A, right), where a dramatic drop in TR is observed after reconfiguration for input frequencies within the initial passband, demonstrating the self-tuning behavior to prohibit the wave. Additionally, we present the steady-state response amplitude visualization of the PM under initial position and after self-tuning for a testing frequency of $f = 125 \text{ Hz}$ in Fig. 3B, top, where a wave attenuation along the propagating direction is observed after the reconfiguration. The real-time responses of one input module and one output module (Fig. 3B, bottom) further confirm that the wave is blocked. In addition to the wave blocking task, other frequency-selective wave control tasks, such as the self-adaptive wave passing task to allow the incoming wave to pass through the PM regardless of its input frequency, can be efficiently pursued with the same structure by simply setting a different targeted mapping and training the system for a new set of readout weights (see SI Appendix, Section S4 for the wave passing task).

In addition to being frequency-selective, it is also feasible to achieve amplitude-selective wave propagation control with the same structure but different targets and training. In this case, we can design an amplitude-dependent mapping $X_T = T(A_0)$ between the input amplitudes and targeted actuator displacements by fixing the input frequency. This allows the PM to exhibit different actuation behavior based on input amplitudes. For example, we can create an unconventional amplitude-selective wave filter that blocks wave with moderate-amplitude incoming waves while allowing the propagation of low and high-amplitude incoming waves (see Fig. 3C and Fig. 3D). In this scenario, we set the input frequency to $f = 125$ Hz and vary the input amplitude by adjusting the excitation voltage amplitudes of the waveform generator. We excite the PM from the initially undeformed configuration and collect 100 sets of wave dynamics data with excitation voltage amplitudes uniformly drawn from $0.1$ V to $10$ V, which are randomly split into 80 training samples and 20 testing samples. With this setup, at the initial position, the incoming wave can propagate along the PM regardless of its input amplitude. Subsequently, the PM reservoir is trained to learn a targeted mapping that filters (suppresses) the incoming waves with moderate amplitudes ($3.33$ V ~ $6.67$ V). The results in Fig. 3 C demonstrate the successful performance of the PM reservoir in this amplitude-selection task, where all the testing samples with wave amplitudes between $3.33$ V and $6.67$ V are effectively suppressed through self-tuning of the PM from passband to stopband. More case studies of amplitude-selection wave control tasks can be found in SI Appendix, Section S5. Note that, within the considered input amplitude range, our system remains small-amplitude vibrations around the equilibria, with a maximal displacement amplitude less than $1 \times 10^{-4}$ m. Hence, we can assume the same theoretical band structure obtained from the linear dispersion analysis.

From the above experimental studies, we have demonstrated that the PRC framework enables the proposed PM to learn from its wave dynamics, perceive the input signal features, make input-dependent decisions, and command corresponding actions for self-adaptive wave functionalities. In essence, we construct and integrate the various MI aspects, including perception, decision-making, and commanding capabilities, in and through the mechanical domain of proposed PM. This innovative approach harnesses the physical computing power inherent in mechanical structures, eliminating the need for bulky additional signal processors or external digital controllers to assess excitation features or guide the linear actuator. The versatility and multi-tasking capabilities of the PRC-based MI are also showcased through diverse studies, achieving both the frequency and amplitude-dependent wave functionality adaptation using the same structure. Such multi-faceted and engineering-relevant MI will contribute to advancing a wide range of wave-based applications, such as vibration suppression, waveguides, lenses, and energy harvesting devices, for more embodied structural autonomy and intelligence.

It is worth noting that with the PRC framework, while simple additive circuits are required to physically implement the linear summation of the readouts, the main aspects of the operational condition perception, decision-making, and commanding are embodied within the dynamics of the PM. To further minimize the reliance on electronics for linear summation, we employ the LASSO regression algorithm for variable selection [36], which help identify the most informative readouts to train the system within the constraint of limited number of readouts. This technique also allow us to reduce the overfitting phenomenon by eliminating redundant or irrelevant readouts. For instance, in the case of the wave blocking task, we select 15 readouts out of total 22 pairs of bent beams to optimize the reservoir testing performance. Through additional numerical investigations, we also discover that by reducing the system noise level we can decrease the number of required readouts to as few as 5, while still ensuring satisfactory reservoir performance (see SI Appendix, Section S6 for details on readout selections).

**Wave-based logic gates.** Building upon the above framework, we further expand the intelligent functionalities to beyond wave isolation and passing. One example is to harness the PM reservoir architecture to construct intelligent wave-based logic gates via the proposed self-adaptive bandgap tuning mechanism. The development of the wave-based logic gates enables the proposed PM to process information carried by elastic waves and make Boolean logic decisions on whether passing or filtering the wave, which can facilitate intelligent and reconfigurable phononic wave-information communication and computing systems. Typically, the realization of the mechano-logic involves the abstraction of mechanical bits, often achieved through multi-stability, dynamical phase transitions or wave transmissions [3]-[10]. However, a challenge in current mechano-logic studies is the ability to realize all basic logic gates, which serve as the building blocks for more complex operations. Our approach based on PRC offers an efficient alternative for creating multiple logic gates within a single PM without altering its design, which is achieved by abstracting its wave transmission characteristics as binary mechanical bits. The binary outputs of the logic gates are defined based on the TR, with low TR representing '0' state (gate 'OFF' and wave blocking) and high TR representing '1' state (gate 'ON' and wave propagating). Single-input logic gates are experimentally realized with the proposed PM based on the amplitude-selective wave propagation (see Fig. 4A-B). The binary inputs to the logic gate are defined based on the input amplitudes, where amplitudes smaller than $5V$ represent $'0'$ bit and amplitudes between $5V$ to $10V$ represent $'1'$ bit. Here, we consider an operation frequency at $f = 125\,\text{Hz}$ for illustration. As an example, Buffer gate can be created by training the PM to block small-amplitude waves and pass large-amplitude ones, resulting in an 'OFF' state for '0' input and an 'ON' state for '1' input. Similarly, a NOT gate is realized by passing small-amplitude waves and blocking large-amplitude waves (see SI Appendix, Section S7 for details on the realization of single-input logic gates). Fig. 4A, left and Fig. 4B, left visualize the successful implementation of the Buffer and NOT gates under a testing amplitude of $1\,V$ ('0' bit) and $7.5\,V$ ('1' bit), with the 'OFF' and 'ON' states clearly distinguishable through wave blocking or passing. Fig. 4A, right and Fig. 4B, right present the average TRs for both training and testing samples. A minimum factor of 5 is observed between the average TR ratios of the 'ON' and 'OFF' states, indicating a significant state difference necessary for effective binary bit abstraction from wave dynamics.

To realize the two-input logic gates, two identical PMs can be assembled in parallel, as depicted in Fig. 4C. The left ends of both PMs are fixed to the same position and their first module on the left are independently excited with two incoming waves, A and B, while their right ends are connected to the same linear motion actuator. We fix the input frequencies of both inputs as previous, and apply the same definition for the binary logic inputs based on their amplitudes. As the two PMs always maintain the identical configurations and wave transmissibility, the output states of the two-input logic gates can be defined based on their average TR. In this case, the dynamics of both PMs serves as a physical reservoir to compute the desired actuator displacements for the corresponding gate outputs. Through numerical simulations, all six basic two-input logic gates (AND, OR, NAND, NOR, XOR and XNOR) are successfully realized within the same double PM reservoir (see SI Appendix, Section S7 for details on the realization of two-input logic gates). Figure 4D demonstrates the testing performance for the six logic gates in term of their average transmission ratios under different combinations of inputs A and B, which clearly indicate the successful realization, with an average TR above 1 for 'ON' states and below 0.05 for 'OFF' states. Thus, we have shown that the PM reservoir can perceive and process the information carried by the input waves, make decisions and command transmitting or filtering the wave-information for corresponding wave-logic outputs. The PRC-based approach efficiently realizes all single-input or two-input logic gates within the same PM system without requiring changes to the structural design. This can serve as a foundation for constructing more sophisticated elastic wave-information networks in higher-dimensional systems and open avenues for future intelligent and reconfigurable phononic communication systems and wave-based computing devices.

**Discussion**

In summary, we have uncovered and investigated an adaptive structure concept with integrated multi-functional MI through an efficient and effective physical computing framework and investigated its efficacy in and through a phononic metastructure (PM). By harnessing the PRC power hidden in the physical neural network dynamics of a tunable PM, we embed and integrate various elements of intelligence (namely perceiving information from sensory processing, making decisions, and commanding actions to actuator) in the mechanical domain of the PM, achieving intelligent wave-physics adaptation. We realize self-adaptive wave transmissions by training our system to tune itself according to different input excitation. Furthermore, we harness the wave manipulation functions to explore wave-based mechano-logic and successfully achieve all the basic logic gates by abstracting the wave transmission characteristics as mechanical bits. Building on these findings on the PM platform, our approach can be expanded to a large number of other high-degree-of-freedom and nonlinear structures, such as origami structures, tensegrity structures and soft robotics [20]–[23], which would have broad impact in various engineering fields. Overall, this research is transformative in offering a systematic foundation towards multi-faceted and integrated mechano-intelligence and in breaking new paths for adaptive structures and material systems.

**Materials and Methods**

Materials and fabrication methods of the experimental prototype are summarized in SI Appendix, section S1. The mathematical modelling of the PM module is introduced in SI Appendix, section S2. Section S2 and S3 detail the derivation of the governing equations and the theoretical band structures via linear dispersion analysis. SI Appendix, section S6 introduces the feature selection method via LASSO regression. Section S4, S5 and S7 present the implementation method of different intelligent wave control tasks.

**Acknowledgments**

The authors acknowledge the support of the University of Michigan Collegiate Professorship. This research is also partially supported by the Automotive Research Center in accordance with Cooperative Agreement W56HZV-19-2-0001 U.S. Army Ground Vehicle System Center in Warren, MI.

**Figures**

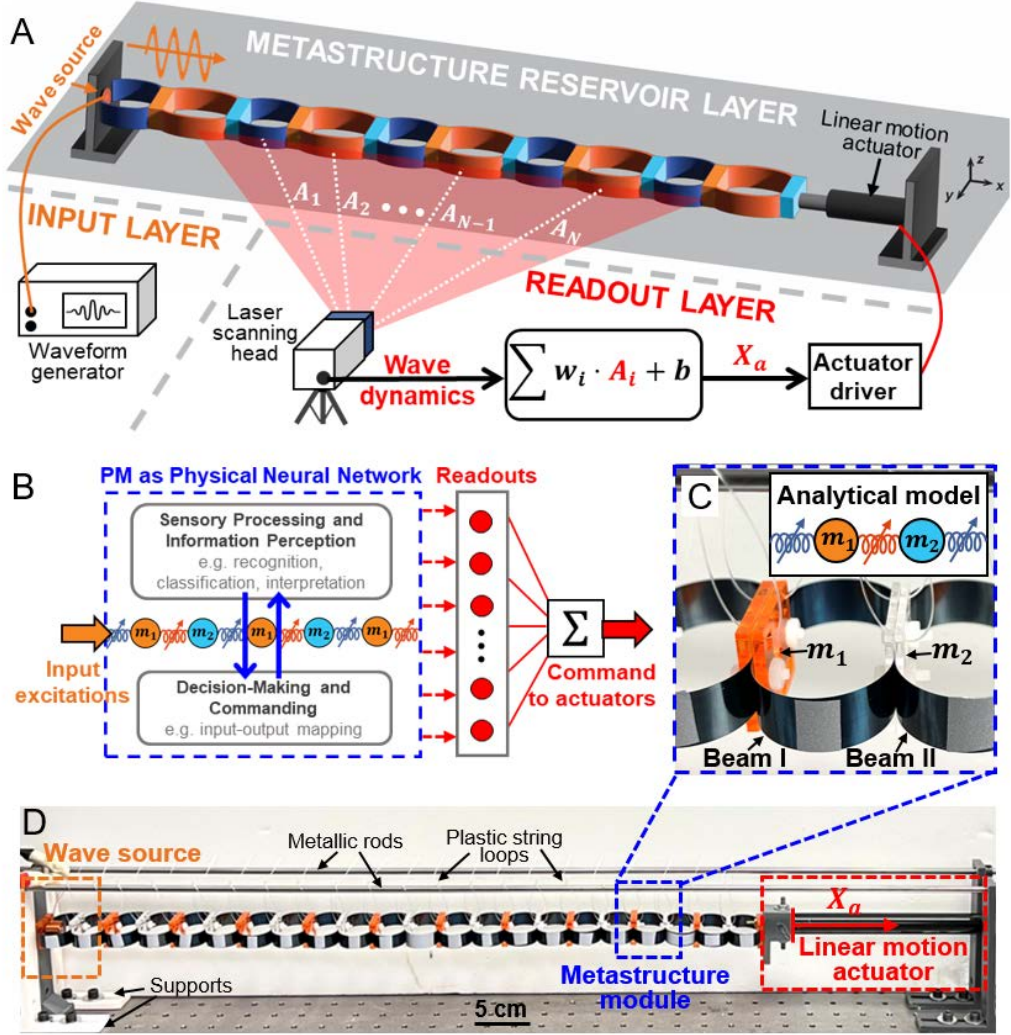

**Figure 1. Mechanically intelligent phononic metastructure.** (A) Conceptual visualization of the PM reservoir setup for intelligence wave propagation control. It is a three-layer architecture, consisting of the input layer to generate input waves, the reservoir layer to provide high-degree-of-freedom wave dynamics as computational resource and the readout layer to compute and output the action command to guide the linear motion actuator. (B) PM as a physical neural network. Various MI aspects, namely perceiving information from sensory processing, making decisions, and commanding actions to actuator, are embedded and integrated in the mechanical domain. (C) A metastructure module and the analytical model via a simplified lumped mass-nonlinear-spring model (inset). (D) Experimental prototype of the phononic metastructure, which are hung with plastic string loops under two guiding rods.

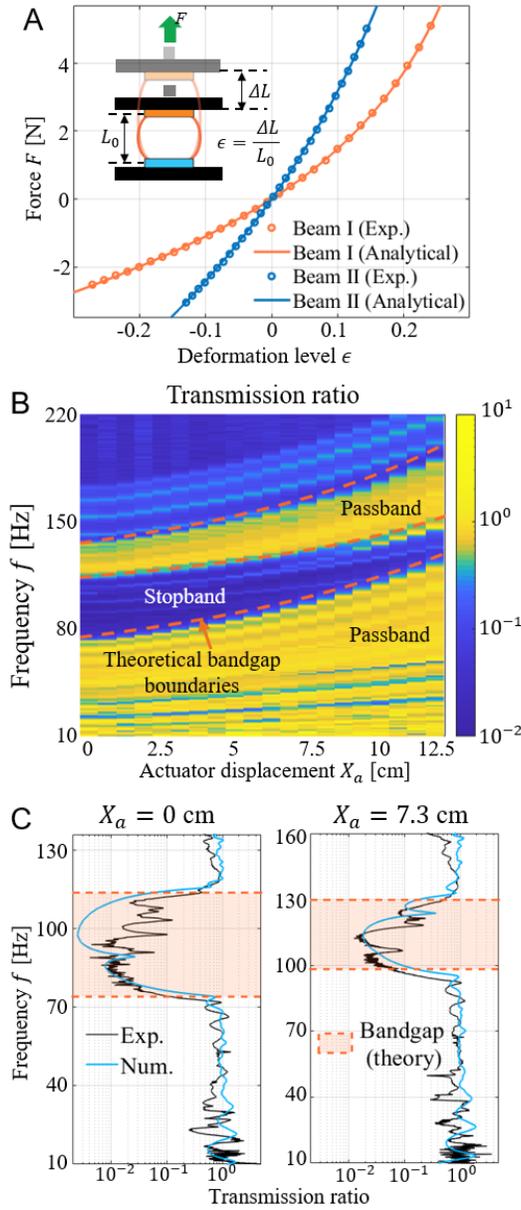

**Figure 2. Tunable wave transmission characteristic of the phononic metastructure under different actuator displacement.** (A) Nonlinear force-deformation relationship of two types of bent beam pairs: the Beam I type (orange) connects the two rigid masses within one module and the Beam II type (blue) inter-connects modules. The results reveals the tunability of the PM under different deformation levels. Excellent agreement is observed between experimental measurements (circles) and analytical modeling predictions (solid line). The inset presents how the force is applied and the definition for the deformation level. (B) The experimental transmission ratio spectrum with respect to the input excitation frequencies under different actuator displacements. The orange dashed lines plot the theoretical bandgap boundaries. (C) The transmission ratios of incoming waves with different excitation frequencies under the initial position $X_a = 0$ cm (left) and a global stretch of $X_a = 7.3$ cm (right). The black curves plot the experimental results, and the blue curves are the numerical simulation results. For comparison, the orange shaded areas show the theoretically predicted bandgaps.

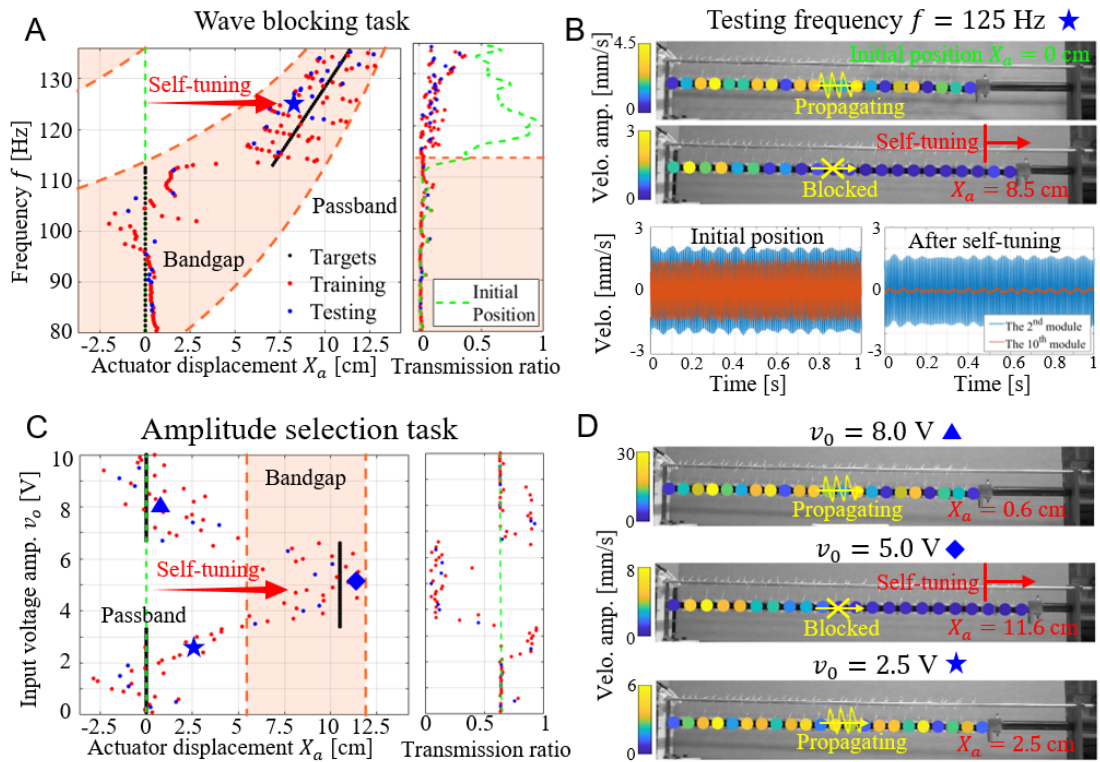

**Figure 3. Self-adaptive wave transmission tasks.** (A) The wave blocking task to prohibit all incoming wave transmissions. The left plot shows the training and testing results, where the green dashed line is the initial position at $X_a = 0$ cm, the black dots are the targeted mapping between the input frequencies and the actuator displacements and the red and blue dots present the training and testing output, respectively. The right plot presents the transmission ratio at the initial condition and the new configuration after performing the task. (B) Visualization of the wave propagation at testing frequency $f = 125$ Hz (top). Both the initial configuration and new configuration after the self-tuning are shown. The real-time response of one input module and one output module (bottom). The orange curves plot the steady-state velocity response of the Beam I at the 2$^{nd}$ module and the blue curves plots that of the 10$^{th}$ module. (C) The amplitude-selection task to pass small and large-amplitude waves and block the mid-amplitude ones. (D) Visualization of the wave propagation under the new configurations after learning the task for three testing input signals with voltage amplitudes of 2.5 V, 5.0 V and 8.0 V. Only for the case with moderate amplitude $v_0 = 8.0$ V, the incoming wave is blocked.

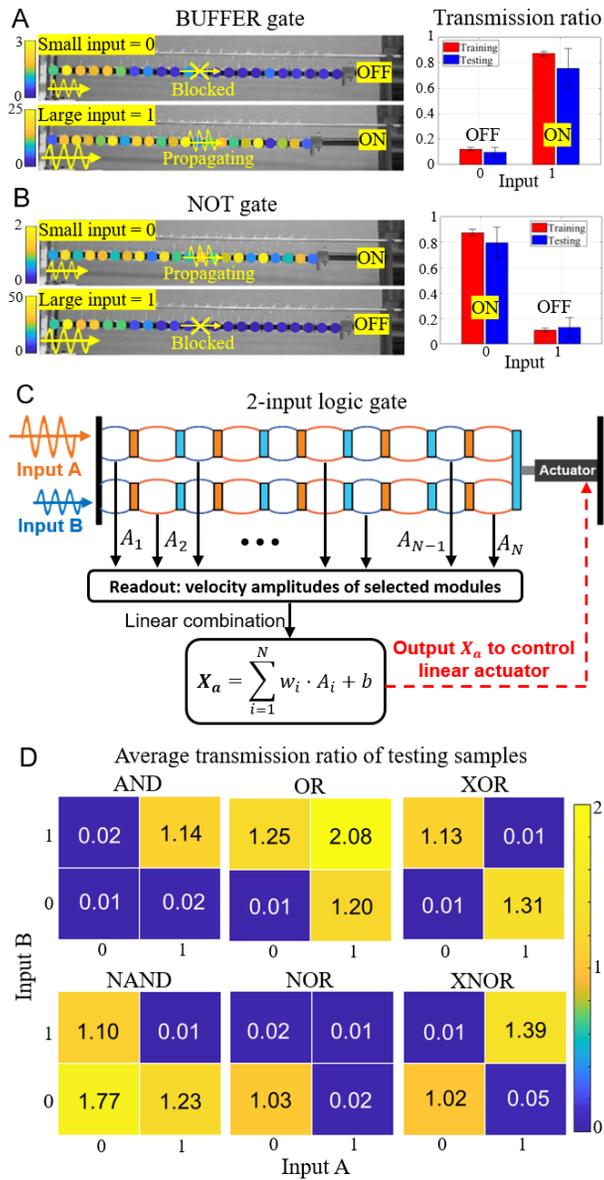

**Figure 4. Wave-based logic gates.** (A) BUFFER gate. The binary inputs are defined by the input amplitude, small-amplitude input as ′0′ and large-amplitude input as ′0′, and the logic outputs are defined by the wave transmission ratio characteristics with the wave blocked being the 'OFF' state and the wave passed being the 'ON' state. The left plot shows the average transmission ratios (with the 95% confidence interval) among the training samples (red bar) and the testing samples (blue bar). (B) NOT gate. All the single-input logic gates are realized in experiments. (C) The schematic to realize the two-input logic gates by assembling two PMs in parallel. (D) Numerical realization of all the two-input logic gates: AND, NAND, OR, NOR, XOR and XNOR, showing the average transmission ratio for each logic gate.

15